\documentclass[twocolumn,showpacs]{revtex4}
\usepackage{graphicx}
\usepackage{dcolumn}
\usepackage{amsmath}

\begin{document}

\preprint{QED Ferromagnetism}

\title[Short Title]{Radiation Induced Landau-Lifshitz-Gilbert 
Damping in Ferromagnets}

\author{S. Sivasubramanian and A. Widom}
\affiliation{Physics Department, Northeastern University, Boston MA, USA} 
\author{Y.N. Srivastava}
\affiliation{Physics Department \& INFN, University of Perugia, Perugia, Italy}

\begin{abstract}
The  Landau-Lifshitz-Gilbert damping coefficient employed in the analysis 
of spin wave ferromagnetic resonance is related to the electrical 
conductivity of the sample. The changing magnetization (with time) 
radiates electromagnetic fields. The electromagnetic energy is then 
absorbed by the sample and the resulting heating effect describes 
magnetic dissipative damping. The ferromagnetic resonance relaxation 
rate theoretically depends on the geometry (shape and size) 
of the sample as well as temperature in agreement with experiment.  
\end{abstract}

\pacs{75.45.+j, 75.10.-b, 75.10.Jm, 75.50.Ww}

\maketitle

The description of ferromagnetic resonance damping in single 
domain samples is conventionally described by a transport 
coefficient \begin{math} {\sf R}  \end{math} in the 
Landau-Lifshitz-Gilbert dynamical equation of motion\cite{1,2}. 
In detail, if \begin{math} {\cal U}[{\bf M},S] \end{math} 
denotes the energy of the domain as a functional of magnetization 
and entropy, then the equation of motion is given by  
\begin{equation}
\frac{\partial {\bf M}}{\partial t}=\gamma  ({\bf M\times H}_{eff}), 
\end{equation}
where the effective magnetic intensity 
\begin{equation}
{\bf H}_{eff}=\left({\bf H}-\frac{\delta {\cal U}}{\delta {\bf M}}\right)
+{\sf \tau }\cdot \left(\frac{\partial {\bf M}}{\partial t}\right)
\end{equation}
and \begin{math} \gamma =(eg/2mc)=(1+\kappa )(e/mc) \end{math} is the 
electronic gyromagnetic ratio. Experimental data are usually expressed 
in terms of the dimensionless tensor transport coefficient  
\begin{math} {\bf \alpha }  \end{math}; i.e.   
\begin{equation}
{\sf \alpha }=|\gamma {\bf M}|{\sf \tau }.
\end{equation}

The transport (tensor) coefficient 
\begin{math} {\sf \tau } \end{math} enters into the Landau-Lifshitz-Gilbert 
equation simply as an experimental parameter\cite{3,4,5,6,7,8,9,10,11}. 
In spite of considerable effort, presently there is 
no generally accepted {\em theoretical} 
picture for the physical source of the irreversibility in ferromagnetic 
resonance. In simple magnon decay models\cite{12}, the ferromagnetic 
relaxation takes place in two stages. A magnon with virtually zero 
wave vector decays into one or more higher wave vector magnons. The 
resulting burst of magnetic energy is later distributed to phonon 
lattice vibrations\cite{13,14,15}. Phonon modulations of the 
dipole-dipole and spin-orbit magnetic anisotropy were among 
the earliest of the relaxation mechanisms\cite{16}. However, 
phonon mechanisms in clean crystals yielded magnetization decay times 
which were much too long when compared with experiment. The situation 
was partially remedied by the notion that the magnons could decay 
via the final density of states as determined by lattice 
imperfections\cite{17,18}. But the only somewhat shorter imperfect 
lattice relaxation times no longer had a convincingly correct 
experimental temperature dependence\cite{19}. Also, the overall relaxation 
always remained longer than experiment. Furthermore, none of the above 
efforts in understanding  \begin{math} {\sf \tau }  \end{math} take note of 
the experimental frequency dependence of damping widths in ferromagnetic 
resonance. There exists (in fact) an ``impedance'' 
\begin{math} {\sf Z}(\zeta ) \end{math} as an 
analytic function of complex frequency (for  
\begin{math} {\Im }m\zeta >0 \end{math}) whose real part  
\begin{equation}
{\sf \tau }(\omega )={\Re }e\left\{{\sf Z}(\omega +i0^+)\right\} 
\end{equation}
describes dissipation.

The purpose of this work is to provide a simple formula for the 
Landau-Lifshitz-Gilbert impedance 
\begin{math} {\sf Z}(\zeta ) \end{math} 
in terms of (frequency dependent) electrical conductivity 
\begin{math} \sigma (\zeta )\end{math}
of the ferromagnetic sample. The final result is that 
\begin{equation}
{\sf Z}(\zeta )=\left(\frac {\sigma (\zeta )}{c^2}\right)
\left<{\sf 1}|{\bf L}|^2-{\bf L}{\bf L}\right>,
\end{equation}
where the spatial average over the sample volume 
\begin{math} \Omega  \end{math} is defined as 
\begin{math} <...>=\Omega^{-1}\int_\Omega (...)d^3 {\bf r} \end{math} 
and the length scale \begin{math} {\bf L}({\bf r}) \end{math} 
is defined as 
\begin{equation}
{\bf L}({\bf r})=-{\bf grad}\int_\Omega \frac{d^3 {\bf r}^\prime }
{|{\bf R}|}\ \ {\rm where}\ \ {\bf R}= {\bf r}- {\bf r}^\prime .
\end{equation}  
The derivation of our central Eqs.(5) and (6) will be given in 
what follows.

The physical basis of our theory of the Landau-Lifshitz-Gilbert 
impedance is as follows: (i) When a spin wave decays into excitations 
which heat the sample, the current 
\begin{math} {\bf J}=c\ curl{\bf M}  \end{math} 
radiates an electromagnetic field. (ii) The {\em power per unit volume}  
absorbed from the radiation by the sample via (say) simple frequency 
independent Ohmic heating would then be 
\begin{equation}
P=\left<\sigma |{\bf E}|^2\right>.
\end{equation} 
(iii) The eddy currents (produced by the 
electric field \begin{math} {\bf E}  \end{math} 
via conductivity) describe the heating 
mechanism for ferromagnetic resonance relaxation 
no matter what the excitation products. 
The specific excitations (for example phonon 
excitations with lattice impurities) are all made manifest via  
the conductivity \begin{math} \sigma  \end{math}. 

In detail, consider the magnetic field \begin{math} {\bf B}  \end{math} 
produced by the magnetization \begin{math} {\bf M} \end{math} 
in the magnetostatic limit; i.e. with 
\begin{equation}
{\bf B}=curl \int_\Omega \left(\frac{\bf M\times R}{|{\bf R}|^3}\right)
d^3{\bf r}^\prime  .
\end{equation}
If one now applies Faraday's law, 
\begin{equation}
c\ curl{\bf E}=-\dot{\bf B},
\end{equation}
to situations in which the magnetization varies slowly in time, 
then one finds a radiated electric field given by Eqs.(8) and (9) 
as  
\begin{equation}
c{\bf E}=\int_\Omega \left(\frac{{\bf R}\times \dot{\bf M}}
{|{\bf R}|^3}\right)d^3{\bf r}^\prime  .
\end{equation}
If the magnetization is uniform in space within the 
ferromagnetic sample, then Eqs.(6) and (10) imply 
\begin{equation}
c{\bf E}({\bf r},t)={\bf L}({\bf r})\times \dot{\bf M}(t).
\end{equation}  
We note in passing that the internal electric field Eq.(11) is closely 
connected to the demagnetization field intensity equation 
\begin{equation}
{\bf H}_d({\bf r},t)=-4\pi {\sf N}({\bf r})\cdot {\bf M}(t) 
\end{equation}
for which the tensor 
\begin{equation}
4\pi {\sf N}=-{\bf \nabla \nabla }
\int \frac{d^3{\bf r}^\prime }{|{\bf R}|}={\bf \nabla L}. 
\end{equation}
It follows that 
\begin{math} Tr{\sf N}=div{\bf L}/(4\pi )=1 \end{math}. 
For a uniformly magnetized sample, Eq.(11) plays the same role 
for the internal electric field that Eq.(12) plays for the internal 
magnetic intensity. The power dissipation per unit volume in 
Eq.(7) may be evaluated using Eq.(11) yielding 
\begin{equation}
P=\frac{1}{c^2}
\left<\sigma \left|{\bf L}\times \dot{\bf M}\right|^2\right>=
\dot{\bf M}\cdot {\sf \tau }\cdot \dot{\bf M} 
\end{equation}
where \begin{math} {\sf \tau}  \end{math} is the 
Landau-Lifshitz-Gilbert transport coefficient tensor entering 
into Eqs.(1) and (2). Eq.(14) implies that 
\begin{equation}
 {\sf \tau }=\frac{\sigma }{c^2}
\left<{\sf 1}|{\bf L}|^2-{\bf L}{\bf L}\right>.
\end{equation}
The above Eq.(15) holds for the case in which 
\begin{math} \sigma  \end{math} and 
\begin{math}  {\sf \tau } \end{math} 
do not depend on frequency. If the conductivity 
depends on frequency \begin{math} \sigma( \zeta ) \end{math}, 
then Eq.(15) is easily generalized to Eqs.(4) and (5) 
which are the central results of this work. Note that the 
tensor nature of \begin{math} {\sf Z}(\zeta )  \end{math} 
implies that ferromagnetic damping depends on the shape and size 
of the sample. This theoretical consequence of our theory is of 
experimental importance\cite{20}.

For the case of a thin film, the vector 
\begin{math} {\bf L} \end{math} is in the direction of 
the normal unit vector \begin{math} {\bf n} \end{math} 
to the film; i.e. 
\begin{math} {\bf L}=4\pi ({\bf n\cdot r}){\bf n} \end{math}. 
The  Landau-Lifshitz-Gilbert transport coefficient tensor 
for a thin film of thickness \begin{math} d \end{math} is then 
\begin{equation}
{\sf \tau}=\left(\frac{4\pi^2 d^2 \sigma }{3c^2}\right)
({\sf 1}-{\bf nn})\ \ \ ({\rm for \ thin\ fims}).
\end{equation}
In most experiments on ferromagnetic films 
\begin{math} {\bf n}\cdot  {\bf M}=0 \end{math}. 
In the plane of the film, \begin{math} {\sf \tau } \end{math} 
can then be described by an isotropic scalar 
\begin{math} \alpha \end{math}; i.e.  
\begin{math} 
\alpha {\sf 1}_{\perp \perp}=
|\gamma {\bf M}|{\sf \tau }_{\perp \perp} 
\end{math} 
where 
\begin{equation}
\alpha =\left(\frac{4\pi^2 d^2 \sigma }{3c^2}\right)
|\gamma {\bf M}|.
\end{equation}
In terms vacuum impedance  
\begin{math} R_{vac}=(4\pi /c) \end{math} 
and the film resistance ``per square''
\begin{math} R\fbox{}=1/(\sigma d) \end{math}, 
one obtains the simple result 
\begin{equation}
{\sf \alpha }=\frac{\pi}{3}\left(\frac{R_{vac}}{R\fbox{}}\right)
\left(\frac{|\gamma {\bf M}|d}{c}\right).
\end{equation}
Since \begin{math} R\fbox{} \end{math} increases with increasing 
temperature\cite{21}, one expects that 
\begin{math} \alpha  \end{math} should fall with increasing 
temperature. This is (in fact) an observed\cite{22,23,24,25} result. 

We have presented above, a physical picture of how the 
Landau-Lifshitz-Gilbert damping of ferromagnetic resonance takes 
place. Just as a static magnetization \begin{math} {\bf M} \end{math} 
produces a demagnetizing magnetic field intensity 
\begin{math} {\bf H}_d \end{math}, a changing magnetization 
\begin{math} \dot{\bf M} \end{math} produces an electric field 
\begin{math} {\bf E} \end{math}. The resulting Ohms law conduction 
current \begin{math} {\bf j}=\sigma {\bf E} \end{math} produces the 
Landau-Lifshitz-Gilbert damping magnetic field intensity 
 \begin{math} {\bf h}=-{\sf \tau }\cdot \dot{\bf M} \end{math}. 
Thus the damping parameter \begin{math} \alpha  \end{math} is 
directly and simply related to the conductivity 
\begin{math} \sigma  \end{math} of the sample. For thin films, 
this relationship is given in Eq.(18). Previously puzzling temperature 
dependences of \begin{math} \alpha  \end{math} appear now as self 
evident.


\begin{thebibliography}{99}

\bibitem{1} L. Landau and L. Lifshitz, {\it Phys. Zeit. Sowjetunion} 
{\bf 8}, 153 (1935).
\bibitem{2} T. A. Gilbert, {\it Armor Research Foundation Rep. No. 11}
Chicago, IL. (1955).
\bibitem{3} C.J. Garcia-Cervera and E. Weinan, 
{\it J. Appl. Phys.} {\bf 90}, 370 (2001).
\bibitem{4} G. Brown, M.A. Novotnyand and P.A. Rikvold, 
{\it J. Appl. Phys.} {\bf 89}, 7588 (2001).
\bibitem{5} K.J. Lee, N.Y. Park and T.D. Lee,  
{\it J. Appl. Phys.} {\bf 89}, 7460 (2001).
\bibitem{6} V.L. Safanov and H.N. Bertram, 
{\it J. Appl. Phys.} {\bf 87}, 5508 (2000).
\bibitem{7} S.E. Russek, S. Kaka and M.J. Donahue, 
{\it J. Appl. Phys.} {\bf 87}, 7070 (2000).
\bibitem{8} J.Z. Sun, {\it Phys. Rev.} {\bf B 62}, 570 (2000).
\bibitem{9} E.D. Boerner, H.N. Bertram and H. Suhl, 
{\it J. Appl. Phys.} {\bf 87}, 5389 (2000).
\bibitem{10} L. He and W.D. Doyle, 
{\it J. Appl. Phys.} {\bf 79}, 6489 (1996).
\bibitem{11} D.O. Smith, {\it J. Appl. Phys.} {\bf 29}, 264 (1958).
\bibitem{12} C. Kittel and E. Abrahams, 
{\it Rev. Mod. Phys.} {\bf 25}, 233 (1953).
\bibitem{13} H. Suhl, {\it IEEE Trans. Magn.} {\bf 34}, 1834 (1998).
\bibitem{14} R.C. Fletcher, R.C. Lecraw and E.G. Spencer, 
{\it Phys. Rev.} {\bf 117}, 955 (1960).
\bibitem{15} H.B. Callen, {\it J. Phys. Chem. Solids} {\bf 4}, 156 (1958). 
\bibitem{16} J.H. Van Vleck, {\it Phys. Rev.}  {\bf 52}, 1178 (1937).
\bibitem{17} A.M. Clogston, H. Suhl, L.R. Walker and P.W. Anderson, 
{\it Phys. Rev.} {\bf 101}, 903 (1956). 
\bibitem{18} V.L. Safanov and H.N. Bertram, 
{\it Phys. Rev.} {\bf B 61}, R14893 (2000). 
\bibitem{19} J.F. Cochran, J.M. Rudd, W.B. Muar, G. Traylang 
and B. Heinrich, {\it J. Appl. Phys.} {\bf 70}, 6545 (1991).
\bibitem{20} D.L. Beke, {\it Cryst. Technol.} {\bf 33}, 1039 (1998).
\bibitem{21} S. P. Lee, C. K. Kim, K. Nahm, M. Mittag, Y. H. Jeong 
and C. Ryu, {\it J. Appl. Phys.} {\bf 81}, 2454 (1997). 
\bibitem{22} S.R. Julian, C.Pfleiderer, E.M. Grosche, N.D. Mathur, 
G.J. Mcmullan, A.J. Diver, I. R. Walker and G.G. Lonarich 
{\it J. Phys.Condens. Matter} {\bf 8}, 9675 (1996).
\bibitem{23} V. Yu Irknin and M.I. Katsnel'son, 
{\it Physics-Uspekhi} {\bf 37}, 659 (1994). 
\bibitem{24} S.U. Jen  and Y. D. Yao, {\it J. Appl. Phys. } {\bf 61}, 
4252 (1987).
\bibitem{25} I. Klik, {\it J. Appl. Phys.} {\bf 73}, 6725 (1993).





\end{thebibliography}
\end{document}